# Magnetoelectric-field helicities and reactive power flows

E. O. Kamenetskii, M. Berezin, and R. Shavit

Microwave Magnetic Laboratory
Department of Electrical and Computer Engineering
Ben Gurion University of the Negev, Beer Sheva, Israel

January 25, 2015

**Abstract**
The dual symmetry between the electric and magnetic fields underlies Maxwell's electrodynamics. Due to this symmetry one can describe topological properties of an electromagnetic field in free space and obtain the conservation law of optical (electromagnetic) helicity. What kind of the field helicity one can expect to see when the electromagnetic-field symmetry is broken? The near fields originated from small ferrite particles with magnetic-dipolar-mode (MDM) oscillations are the fields with the electric and magnetic components, but with broken dual (electric-magnetic) symmetry. These fields – called magnetoelectric (ME) fields – have topological properties different from such properties of electromagnetic fields. The helicity states of ME fields are topologically protected quantum-like states. In this paper, we study the helicity properties of ME fields. We analyze conservation laws of the ME-field helicity and show that the helicity density is related to an imaginary part of the complex power-flow density. We show also that the helicity of ME fields can be a complex value.

**I. Introduction**

Symmetry principles play an important role with respect to the laws of nature. To put into a symmetrical shape the equations, coupling together the electric and magnetic fields, Maxwell added an electric displacement current. Such an additive, introduced for reasons of symmetry, resulted in appearing a unified-field structure: the electromagnetic field. The electric displacement current in Maxwell equations allows correct prediction of magnetic fields in regions where no free current flows and prediction of wave propagation of electromagnetic fields. The dual symmetry between electric and magnetic fields underlies the conservation of energy and momentum for electromagnetic fields [1]. It can be connected also with conservation of polarization of the electromagnetic field. In particular, this symmetry underlies the conservation of optical (electromagnetic) helicity [2 – 4]. As it is stated in Ref [4], the dual electromagnetic theory inherently contains straightforward and physically meaningful descriptions of the helicity, spin and orbital characteristics of light.

    What kind of the source-free time-varying field structure one can expect to see when an electric displacement current is neglected and so the electromagnetic-field symmetry (dual symmetry) of Maxwell equations is broken? As one of examples of such a symmetry breaking, we can refer to the field structures studied in non-conductive artificial electromagnetic materials that exhibit zero (or near-zero) permittivity [5, 6]. For these materials, no Maxwell correction (no electric displacement

current) exists and the fields are described by three differential equations (instead of the four-Maxwell-equation description of electromagnetic fields):

$$\nabla \cdot \vec{B} = 0, \tag{1}$$

$$\vec{\nabla} \times \vec{E} = -\frac{\partial \vec{B}}{\partial t}, \tag{2}$$

$$\vec{\nabla} \times \vec{H} = 0. \tag{3}$$

In an assumption that a zero-permittivity medium is magnetically isotropic, from Eqs. (2), (3) it follows that $\vec{\nabla}^2 \vec{E} = 0$. One has the static-like fields. Light passing through such a material experiences no phase shift. Evidently, no unified-field retardation effects can be observed in structures created by the materials described in Refs [5, 6].

It appears, however, that even without the Maxwell's displacement current, certain retardation-effect fields, described by Eqs. (1) – (3), can exist. Such fields, lacking a dual electric-magnetic symmetry, are exhibited in a small sample of a dielectric medium with strong temporal dispersion of magnetic susceptibility – a ferromagnetic-resonant medium. For harmonic fields in this medium, an averaged density of the electromagnetic energy is expressed as $\bar{U} = \frac{1}{2}\left[\varepsilon E_\alpha E_\alpha^* + \frac{\partial(\omega \mu_{\alpha\beta})}{\partial \omega} H_\alpha H_\beta^*\right]$, where $\varepsilon$ is the medium scalar permittivity and $\mu_{\alpha\beta}$ are the components of the permeability tensor $\ddot{\mu}$. In a small ferrite sample (with sizes much less than the free-space electromagnetic wavelength), one has negligibly small variation of electric energy, so that a dynamical process in a sample is described by three differential equations (1) – (3) [7]. Contrary to non-magnetic structures with zero permittivity [5, 6], in this case one can observe the unified-field (with coupled electric and magnetic field components) retardation effects. These oscillations in small ferrite particles are called magnetostatic-wave (MS-wave) or magnetic-dipolar-mode (MDM) oscillations [7 – 10].

MDM oscillations in small ferrite spheres excited by external microwave fields were experimentally observed, for the first time, by White and Solt in 1956 [11]. Afterwards, experiments with disk-form ferrite specimens revealed unique spectra of oscillations. While in a case of a ferrite sphere one observed only a few wide absorption peaks of MDM oscillations [11], for a ferrite disk there was a multiresonance (atomic-like) spectrum with very sharp resonance peaks [12 – 14]. Analytically, it was shown [15, 16] that, contrary to spherical geometry of a ferrite particle analyzed in Ref. [17], the quasi-2D geometry of a ferrite disk gives the Hilbert-space energy-state selection rules for MDM spectra. MDM oscillations in a quasi-2D ferrite disk are macroscopically coherent quantum states, which experience broken mirror symmetry and also broken time-reversal symmetry [18, 19]. There are helical-wavefront oscillations [20 – 24]. Free-space microwave fields, emerging from magnetization dynamics in quasi-2D ferrite disk, carry orbital angular momentums and are characterized by power-flow vortices and non-zero helicity. Symmetry properties of these fields – called magnetoelectric (ME) fields – are different from symmetry properties of free-space electromagnetic (EM) fields. For an incident electromagnetic field, the MDM ferrite disk looks as a trap with focusing to a ring, rather than a point.

ME-coupling properties, observed in the near-field structure, are originated from magnetization dynamics of MDMs in a quasi-2D ferrite disk. In general, ME-coupling effects manifest in numerous macroscopic phenomena in solids. Physics underlying these phenomena becomes evident through a symmetry analysis. In isolating crystal materials, in which both spatial inversion and time-reversal symmetries are broken, a magnetic field can induce electric polarization and, conversely, an electric field can induce magnetization [7, 25]. Without requirements of a special kind of a crystal lattice, a ME-coupling term appears in magnetic systems with topological structures of magnetization. In particular, there can be chiral, toroidal, and vortex structures of magnetization [26, 27]. Other examples on a role of magnetization topology in the ME-coupling effects concern orbital magnetization. As it was discussed in Refs. [28, 29], an adequate description of magnetism in magnetic materials should not only include the spin contribution, but also should account for effects originating in the orbital magnetism. It was shown that in the two-dimensional case, orbital magnetization is exhibited due to exceeding of chiral-edge circulations in one direction over chiral-edge circulations in opposite direction [29]. Recently, it was shown that ME coupling can occur also in isotropic dielectrics due to an effect of orbital ME polarizability – topological ME-coupling effect [30 – 32]. In such a case, one has the contribution of orbital currents to the ME coupling. The orbital ME polarizability is due to the pseudoscalar part of the ME coupling and is equivalent to the addition of a term to the electromagnetic Lagrangian – the axion electrodynamics term [33]. That is why the orbital ME response in isotropic dielectrics is referred as the axion orbital ME polarizability [30, 31].

In this paper, we will show how both the problems of optical (electromagnetic) helicity [2 – 4] and the problems of ME-coupling effects due to magnetization topology [25 – 32] can underlie the helicity properties of microwave ME fields originated from quasi-2D ferrite disks with MDM oscillations. We analyze conservation laws of the ME-field helicity. The paper is organized as follows. Study of helicity of ME fields is preceded by an analysis of helicity in Maxwell's electromagnetism. This analysis is given in Section II of the paper. In Section III, we study general properties of helicity of ME fields. In Section IV, we analyze how the helicity density is related to reactive power flows originated from MDM ferrite disks. In Sections V and VI, we show the numerical studies of complex power flows and complex helicity parameters. Quantized topological characteristics of the ME fields arise from the MS-wave spectral-problem solutions for MDMs in a quasi-2D ferrite disk. In Section VII, we make a brief analytical examination of the ME-field helicities and power flows based on such MS-wave spectral-problem solutions. In Section VIII, we discuss general aspects of the ME-field topology and give a conclusion of our studies shown in the paper.

## II. FIELD HELICITY IN MAXWELL'S ELECTROMAGNETISM

At the beginning of this section, we should, probably, raise the issue: Which term is more relevant for description of the light twistness in free space: "chirality" or "helicity"? There is evident ambiguity in using these terms in literature. Chirality is usually considered as the property related to handedness. In condensed matter physics, chirality is to be associated with enantiomorphic pairs which induce optical activity. At the same time, the wave helicity, related to a Faraday effect, does not require a lack of structural symmetry. In elementary particle physics, helicity represents the projection of the particle spin at the direction of motion. In this case, chirality is considered as the same as the helicity only when the particle mass is zero or it can be neglected. On the other hand, one can be faced with misuse of the term chirality as a synonym of handedness. Simplest examples are a Cartesian coordinate system and a Lorentz force. Both structures are handed. However, based on the known definition "An object is chiral if no mirror image of the object can be superimposed on itself", one can see that a Cartesian coordinate system (three unit polar vectors) is chiral, while a Lorentz force (one axial vector and two polar vectors) is achiral. This discussion shows that there is no definite answer to the above question. We will use both terms relating, mainly, to the literature sources.

Helicity admits topological interpretation in relation to the linkage of vortex lines of the flow. In plasma physics, the helicity of a static magnetic field is considered as a measure of the screwness of the magnetic line and is defined as

$$\mathcal{M} = \int \vec{A} \cdot \nabla \times \vec{A} \, dV ,\qquad(4)$$

where $\vec{A}$ is a vector potential related to the magnetic induction field: $\vec{B} = \nabla \times \vec{A}$. For the static magnetic field, the helicity characterizes to what extent magnetic lines are coupled with each other. For a single magnetic line, the helicity parameter estimates the screwness of this line. It shows the extent to which a magnetic field "wraps around itself" [34 – 36].

Helicity of electromagnetic fields is described by different aspects. All these aspects are related, anyway, to symmetry properties of Maxwell equations. The relativistic generalization of helicity for an arbitrary free-space electromagnetic field is defined as [37, 38]

$$\mathcal{H} = \frac{1}{2} \int \left( \vec{A} \cdot \nabla \times \vec{A} - \vec{C} \cdot \nabla \times \vec{C} \right) dV ,\qquad(5)$$

where $\vec{C}$ is a vector potential related to the electric field: $\vec{E} = \nabla \times \vec{C}$. In such a relativistic generalization, the electromagnetic (optical) helicity is a measure of the screwness of the electromagnetic field. In the quantum electrodynamics representation, it coincides with the difference between the number of the right and left circularly polarized photons composing the electromagnetic field. The electromagnetic helicity $\mathcal{H}$ is a time-even Lorentz pseudoscalar with the dimensions of an angular momentum. It is a conserved quantity in that: $\frac{d\mathcal{H}}{dt} = 0$ [2 – 4, 37].

Maxwell's equations are invariant when the electric and the magnetic fields $\vec{E}$ and $\vec{B}$ mix via a rotation by an arbitrary angle $\xi$, as

$$\begin{pmatrix} \vec{E}' \\ \vec{B}' \end{pmatrix} = \begin{pmatrix} \cos\xi & \sin\xi \\ -\sin\xi & \cos\xi \end{pmatrix} \begin{pmatrix} \vec{E} \\ \vec{B} \end{pmatrix}.\qquad(6)$$

For a real angle $\xi$, this transformation leaves invariant such quadratic forms as the Poynting vector and energy density [1]. The duality rotation (6) generates the same rotation of the vector potentials:

$$\begin{pmatrix} \vec{A}' \\ \vec{C}' \end{pmatrix} = \begin{pmatrix} \cos\xi & \sin\xi \\ -\sin\xi & \cos\xi \end{pmatrix} \begin{pmatrix} \vec{A} \\ \vec{C} \end{pmatrix}.\qquad(7)$$

It means that the electromagnetic helicity density $h = \frac{1}{2}\left(\vec{A}\cdot\nabla\times\vec{A} - \vec{C}\cdot\nabla\times\vec{C}\right)$ retains its form under a duality rotation (7) [3, 4]. The electromagnetic helicity is not the only quantity in electromagnetic theory that describes the angular momentum associated with polarization. Also, one can obtain the spin angular momentum of light. The spin density of the field, $\vec{s} = \frac{1}{2}\left(\vec{E}\times\vec{A} + \vec{B}\times\vec{C}\right)$, has the dimension of an angular momentum per unite volume and retains its form under a duality rotation (7). The helicity density $h$ and the spin density $\vec{s}$ are related by a continuity equation. Similar to electromagnetic helicity, electromagnetic-field spin is a conserved quantity. The symmetries underlying these conservation laws are dual symmetries for the electric and magnetic fields in Maxwell equations [3, 4].

While reference to the vector potentials raises the question on gauge dependence, explicit reference to the fields is gauge invariant. Recently, considerable interest has been aroused by a rediscovered measure of helicity in optical radiation – commonly termed optical chirality density – based on the Lipkin's "zilch" for the fields [39]. The optical chirality density is defined as [39 – 42]:

$$\chi = \frac{\varepsilon_0}{2}\vec{E}\cdot\nabla\times\vec{E} + \frac{1}{2\mu_0}\vec{B}\cdot\nabla\times\vec{B}. \tag{8}$$

The optical chirality density is related to the corresponding chirality flow via the differential conservation law:

$$\frac{\partial \chi}{\partial t} + \vec{\nabla}\cdot\vec{f} = 0, \tag{9}$$

where

$$\vec{f} = \frac{\varepsilon_0 c^2}{2}\left[\vec{E}\times(\nabla\times\vec{B}) - \vec{B}\times(\nabla\times\vec{E})\right]. \tag{10}$$

For time-harmonic fields (with the field time dependence $e^{i\omega t}$), the time-averaged optical chirality density is calculated as [39 – 42]

$$\chi = \frac{\omega\varepsilon_0}{2}\text{Im}\left(\vec{E}^*\cdot\vec{B}\right), \tag{11}$$

where vectors $\vec{E}$ and $\vec{B}$ are complex amplitudes of the electric and magnetic fields. This is a time-even, parity-odd pseudoscalar parameter. Lipkin showed [39], that the chirality density is zero for a linearly polarized plane wave. However, for a circularly polarized wave, Eq. (11) gives a nonvanishing quantity. Moreover, for right- and left-circularly polarized waves one has opposite signs of parameter $\chi$.

The effect of optical chirality was applied recently for experimental detection and characterization of biomolecules [43]. The chiral fields were generated by the optical excitation of plasmonic planar chiral structures. Excitation of molecules is considered as a product of the parameter of optical chirality with the inherent enantiometric properties of the material. In experiments [43], the evanescent near-field modes of plasmonic oscillations are involved. In continuation of these studies, a detailed and systematic numerical analysis of the near-field chirality in different plasmonic nanostractures was made in Ref [44]. However, in connection with the results obtained in Refs. [43, 44], an important question arises: Whether, in general, the expressions (8), (11) are applicable for description of the chiroptical near-field response? The near-field chiroptical properties shown in Refs. [43, 44], are beyond the scope of the Lipkin's analysis, which was made based only on the plane wave consideration.

In a case of Eq. (11), the electric field is parallel to the magnetic field with a time-phase delay of $90°$. In Ref. [45], it was shown that in an electromagnetic standing-wave structure, designed by interference of two counter-propagating circular polarized plane waves with the same amplitudes, there are certain planes where the electric and magnetic fields are collinear with each other and are not time-phase shifted. Such a field structure results in appearance of the energy density expressed as

$$W^{(me)} \propto \frac{1}{2}\text{Re}\left(\vec{E}^* \cdot \vec{B}\right). \tag{12}$$

The authors in Ref. [45] call this energy as the magnetoelectric energy. Intuitively, it was assumed [45] that this ME energy density of plane monochromatic waves can be related to the reactive power flow density (or imaginary Poynting vector) [1]:

$$\vec{S}^{(me)} \propto \frac{1}{2}\text{Im}\left(\vec{E}^* \times \vec{B}\right). \tag{13}$$

In Refs. [21 – 24, 46], it was shown that the ME properties can be observed in the vacuum-region fields originated from ferrite-disk particles with MDM oscillations. Contrary to Refs. [45], there are not the states of propagating-wave fields. There are quantized states of the ME near fields.

### III. HELICITY OF THE ME NEAR FIELDS

Differential Eqs. (1) – (3), together with the constitutive relation

$$\vec{B} = \vec{\mu} \cdot \vec{H} \ , \tag{14}$$

describe the fields in small ferrite particles at the ferromagnetic-resonance frequencies [7 – 10]. However, formal use of these three differential equations, Eqs. (1) – (3), does not allow formulation of the spectral problem for MDM oscillations. Without Eq. (2), using only Eqs. (1) and (3) [and the constitutive relation (14)], one obtains the Walker equation for magnetostatic-potential (MS-potential) wave function $\psi$ (introduced based on a relation $\vec{H} = -\vec{\nabla}\psi$) [17]. For a quasi-2D ferrite disk, the Walker-equation differential operator (together with the homogeneous boundary conditions for function $\psi$ and its derivatives) gives the energy eigenstate spectrum of MDM oscillations [15, 16, 18]. There are so called *G*-mode spectral solutions. When we aim to obtain the MDM spectral solutions taking into account also the electric fields in a ferrite disk, we have to consider the boundary conditions for a magnetic flux density, $\vec{B} = -\vec{\mu} \cdot \vec{\nabla}\psi$. In this case (because of specific boundary conditions on a lateral surface of a ferrite disk), one has the helical-mode resonances and the spectral solutions are described by double-valued functions [18, 19,]. There are so called *L*-mode spectral solutions. For *L* modes, the electric field in a vacuum region near a ferrite disk has two parts: the curl-field component $\vec{E}_c$ and the potential-field component $\vec{E}_p$ [21]. While the curl electric field $\vec{E}_c$ in vacuum we define from the Maxwell equation $\vec{\nabla} \times \vec{E}_c = -\mu_0 \frac{\partial \vec{H}}{\partial t}$, the potential electric field $\vec{E}_p$ in vacuum is calculated by integration over the ferrite-disk region, where the sources (magnetic currents $\vec{j}^{(m)} = \frac{\partial \vec{m}}{\partial t}$) are given. Here $\vec{m}$ is dynamical magnetization in a ferrite disk. It was shown that in vacuum near a ferrite disk, the regions with non-zero scalar product $\vec{E}_p \cdot \left( \vec{\nabla} \times \vec{E}_c \right)$ can exist [21].

In a general form, we introduce a notion of the ME-field helicity density expressed as [21 – 23, 46]:

$$F \propto \vec{E} \cdot \left( \vec{\nabla} \times \vec{E} \right) = \vec{E}_p \cdot \left( \vec{\nabla} \times \vec{E}_c \right). \tag{15}$$

Formally, this parameter can be considered as a parameter $\chi$ described by Eq. (8), but with an additional condition of the magnetostatic ($\vec{\nabla} \times \vec{H} = 0$) description. With such a meaning, we represent the ME-field helicity density as

$$F = \frac{\varepsilon_0}{2} \vec{E} \cdot \nabla \times \vec{E}. \tag{16}$$

The product $\vec{E} \cdot \left( \vec{\nabla} \times \vec{E} \right)$ is a measure of the screwness of the electric field. It is equal to the electric field $\vec{E}$ on the points lying in the screw axis times the vorticity $\vec{\nabla} \times \vec{E}$. As the curl of a vector measures its rotation around a point, the product $\vec{E} \cdot \left( \vec{\nabla} \times \vec{E} \right)$ gives how much $\vec{E}$ rotates around itself times its own modulus. This product evaluates to what degree vector $\vec{E}$ resembles a helix.

For time-harmonic fields ($\propto e^{i\omega t}$), the time-averaged helicity density parameter was calculated in a vacuum near-field region as [21 – 23, 46]:

$$F = \frac{\varepsilon_0}{4} \mathrm{Im} \left\{ \vec{E} \cdot \left( \vec{\nabla} \times \vec{E} \right)^* \right\}. \tag{17}$$

The ME-field helicity density is nonzero only at the resonance frequencies of MDMs. It arises from double-helix resonances of MDM oscillations in a quasi-2D ferrite disk [19]. At the MDM frequency $\omega = \omega_{MDM}$, we have for magnetic induction $\vec{B} = \frac{i}{\omega_{MDM}} \left( \vec{\nabla} \times \vec{E} \right)$. So, Eq. (17) can be rewritten as

$$F = \frac{\omega_{MDM}\varepsilon_0}{4}\operatorname{Im}\left\{i\vec{E}\cdot\vec{B}^*\right\} = \frac{\omega_{MDM}\varepsilon_0}{4}\operatorname{Re}\left\{\vec{E}\cdot\vec{B}^*\right\} = \frac{\omega_{MDM}}{4c^2}\operatorname{Re}\left\{\vec{E}\cdot\vec{H}^*\right\}, \qquad (18)$$

where $c = 1/\sqrt{\varepsilon_0\mu_0}$. From this equation, one can see that the helicity density $F$ transforms as a pseudo-scalar under space reflection $\mathcal{P}$ and it is odd under time reversal $\mathcal{T}$. This is a time-odd, parity-odd pseudoscalar parameter. At the MDM resonances, one observes macroscopically coherent vacuum states near a ferrite disk. These vacuum states of the field experience broken mirror symmetry and also broken time-reversal symmetry. Whenever a pseudo-scalar axion-like field is introduced in the theory, the dual symmetry is spontaneously and explicitly broken [33]. Our studies shown in Refs. [21 – 24, 46] clearly verify the properties of helicity for ME fields originated from MDM resonances in a ferrite disk. Evidently, for regular electromagnetic fields $\operatorname{Re}\left\{\vec{E}\cdot\vec{B}^*\right\} \equiv 0$,

We represent now the potential electric field as $\vec{E}_p = -\vec{\nabla}\phi$, where $\phi$ is an arbitrary electrostatic-potential function. With this representation, we can write:

$$F = \frac{\omega_{MDM}\varepsilon_0}{4}\operatorname{Re}\left\{\vec{E}\cdot\vec{B}^*\right\} = -\frac{\omega_{MDM}\varepsilon_0}{4}\operatorname{Re}\left\{\vec{\nabla}\phi\cdot\vec{B}^*\right\} = -\frac{\omega_{MDM}\varepsilon_0}{4}\left\{\vec{\nabla}\cdot\operatorname{Re}\left(\phi\vec{B}^*\right)\right\}. \qquad (19)$$

Here we took into account that $\vec{\nabla}\cdot\vec{B} = 0$. Based on this equation, one can introduce a quantity of the time-averaged ME-energy density:

$$\vec{\nabla}\cdot\operatorname{Re}\left(\phi\vec{B}^*\right) \equiv -\eta. \qquad (20)$$

The quantity $\phi\vec{B}^*$ can be considered as the time-averaged ME-energy flow. For the helicity density we can write:

$$F = \frac{\omega_{MDM}\varepsilon_0}{4}\eta. \qquad (21)$$

The regions of the positive and negative helicity density [21 – 24, 46] can be described, respectively, as the regions with positive and negative ME-energy density $\eta$. Since the helicity factor $F$ shows what is degree of a twist between the $\vec{E}$ and $\vec{H}$ vectors compared to a regular EM-field configuration (with mutually perpendicular $\vec{E}$ and $\vec{H}$ vectors), the ME energy can be considered as energy of a torsion degree of freedom [21 – 23]. Because of time-reversal symmetry breaking, all the regions with positive helicity become the regions with negative helicity (and vice versa), when one changes a direction of a bias magnetic field:

$$F^{\vec{H}_0\uparrow} = -F^{\vec{H}_0\downarrow}. \qquad (22)$$

This equation can be written also as

$$\eta(-\vec{H}_0) = -\eta(\vec{H}_0). \qquad (23)$$

Let us define the helicity as an integral of the ME-field helicity density over the entire near-field vacuum region of volume $V$ (which excludes a region of a ferrite disk):

$$\mathcal{H} = \int_V F dV = \frac{\omega_{MDM}\varepsilon_0}{4} \int_V \text{Re}\left\{\vec{E}_p \cdot \vec{B}^*\right\} dV = \frac{\omega_{MDM}\varepsilon_0}{4} \int_V \eta dV . \tag{24}$$

The question arises: Whether do we have the "helicity neutrality", i. e. $\mathcal{H} = \frac{\omega_{MDM}\varepsilon_0}{4} \int_V \eta dV = 0$? To answer this question we can rely on the following simple analysis. With use of the transformation

$$\mathcal{H} = -\frac{\omega_{MDM}\varepsilon_0}{4} \int_V \text{Re}\left\{\vec{\nabla} \cdot \left(\phi \vec{B}^*\right)\right\} dV = -\frac{\omega_{MDM}\varepsilon_0}{4} \oint_S \text{Re}\left\{\phi \vec{B}^* \cdot \vec{n}\right\} dS , \tag{25}$$

we can conclude that when the normal component of $\vec{B}$ vanishes at some boundary inside which the fields $\vec{B}$ and $\vec{E}_p$ are confined (i.e. when $\vec{B} \cdot \vec{n} = 0$ at the boundary), the quantity $\mathcal{H}$ is equal to zero. The quantity $\mathcal{H}$ is also equal to zero when the fields are with finite energy and the quantity $\phi\vec{B}^*$ decreases sufficiently fast at infinity.

On the other hand, inside the vacuum-region volume $V$ there are a region $V^{(+)}$ where the helicity is a non-zero positive quantity:

$$\mathcal{H}^{(+)} = \frac{\omega_{MDM}\varepsilon_0}{4} \int_{V^{(+)}} \eta^{(+)} dV > 0 \tag{26}$$

and a region $V^{(-)}$ where the helicity is a non-zero negative quantity:

$$\mathcal{H}^{(-)} = \frac{\omega_{MDM}\varepsilon_0}{4} \int_{V^{(-)}} \eta^{(-)} dV < 0 . \tag{27}$$

For the entire volume $V = V^{(+)} + V^{(-)}$, we have

$$\mathcal{H} = \mathcal{H}^{(+)} + \mathcal{H}^{(-)} = 0 . \tag{28}$$

Also, we should have

$$\left|\mathcal{H}^{(+)}\right| = \left|\mathcal{H}^{(-)}\right| . \tag{29}$$

Such "helicity neutrality" can be considered as a specific conservation law of helicity.

The helicity appears only at the MDM resonances. This quantized quantity of the helicity is represented as

$$\left|\mathcal{H}^{(+)}\right| = \left|\mathcal{H}^{(-)}\right| = \frac{\omega_{MDM}\varepsilon_0}{4}\int_{V^{(+)}}\left|\eta^{(+)}\right|dV = \frac{\omega_{MDM}\varepsilon_0}{4}\int_{V^{(-)}}\left|\eta^{(-)}\right|dV = kn, \tag{30}$$

where $n = 1, 2, 3 \ldots$ is the MDM-resonance number and $k$ is a dimensional coefficient proportionality. Eq. (30) shows also quantization of the positive and negative ME-energy.

## IV. THE HELICITY FACTOR AND REACTIVE POWER FLOW

For time-variation harmonic electromagnetic fields ($\propto e^{i\omega t}$), in the absence of losses and electric-current sources, the imaginary part of the energy balance equation shows that the density of the reactive or stored energy is related to an imaginary part of the complex power-flow density: $2\omega(w_e - w_m) = \frac{c}{8\pi}\vec{\nabla}\cdot\text{Im}\left(\vec{E}\times\vec{H}^*\right)$, where the energy densities $w_e = \frac{c}{16\pi}\left(\vec{E}\cdot\vec{D}^*\right)$ and $w_m = \frac{c}{16\pi}\left(\vec{B}\cdot\vec{H}^*\right)$ are real quantities [1]. By analogy with electromagnetic fields, we will call the vector $\text{Im}\,\vec{E}\times\vec{H}^*$ reactive power flow density. In our case of the MDM ME fields, we have reactive power flows in the near-field vacuum regions, which are different from such flows of regular electromagnetic fields.

In the near-field vacuum area of a quasi-2D ferrite disk with MDM resonances, one has in-plane rotating electric- and magnetic-field vectors localized at a center of a disk [21]. This field structure, shown schematically in Fig. 1, is characterized by the helicity factor. As we will show, for the spinning electric- and magnetic-field vectors, a time averaged real part of a scalar product is related to a time averaged imaginary part of a vector product of the electric and magnetic fields. This will allow making a definite conclusion that for ME fields, the helicity density is related to the reactive power flow density $\text{Im}\,\vec{E}\times\vec{H}^*$. Let us consider the electric- and magnetic-field vectors circularly rotating in the $xy$ plane in a near-field vacuum region at the disk center. Assuming the counterclockwise rotation, we have

$$\vec{E} = a(\hat{\vec{x}} + i\hat{\vec{y}}) \quad \text{and} \quad \vec{H} = b(\hat{\vec{x}} + i\hat{\vec{y}}), \tag{31}$$

where $a = |a|$ and $b = |b|e^{i\vartheta}$ are complex amplitudes, $\hat{\vec{x}}$ and $\hat{\vec{y}}$ are unit vectors of the corresponding vectors along $x$ and $y$ axis, and $\vartheta$ is an arbitrary angle within $0 \leq \vartheta < 90°$. A scalar product of the electric-field vector and the complex conjugate magnetic-field vector gives:

$$\vec{E}\cdot\vec{H}^* = \frac{1}{2}|a||b|e^{-i\vartheta}(\hat{\vec{x}}+i\hat{\vec{y}})\cdot(\hat{\vec{x}}-i\hat{\vec{y}}) = |a||b|[\cos\vartheta - i\sin\vartheta]. \tag{32}$$

We have $\frac{1}{2}\text{Re}\left(\vec{E}\cdot\vec{H}^*\right) = \frac{1}{2}|a||b|\cos\vartheta$. Now consider the magnetic-field vector $\vec{H}'$, which is $90°$ time-shifted with respect to vector $\vec{H}$, that is $\vec{H}' = i\vec{H} = b(-\hat{\vec{y}} + i\hat{\vec{x}})$. For a vector product of the vectors $\vec{E}$ and $\left(\vec{H}'\right)^*$, we have:

$$\vec{E} \times (\vec{H}')^* = \frac{1}{2}|a||b|e^{-i\vartheta}(\hat{\vec{x}} + i\hat{\vec{y}}) \times (-\hat{\vec{y}} - i\hat{\vec{x}}) = -|a||b|\hat{\vec{z}}[\cos\vartheta - i\sin\vartheta], \tag{33}$$

where $\hat{\vec{z}}$ is the unit vector along $z$ axis. At the same time, a vector product of the electric-field vector and the complex conjugate magnetic-field vector $\vec{H}$ gives:

$$\vec{E} \times \vec{H}^* = \frac{1}{2}|a||b|e^{-i\vartheta}(\hat{\vec{x}} + i\hat{\vec{y}}) \times (\hat{\vec{x}} - i\hat{\vec{y}}) = -i|a||b|\hat{\vec{z}}[\cos\vartheta - i\sin\vartheta] = -|a||b|\hat{\vec{z}}[\sin\vartheta + i\cos\vartheta]. \tag{34}$$

Evidently, $\operatorname{Im}\vec{E} \times \vec{H}^* = \operatorname{Re}\vec{E} \times (\vec{H}')^*$. When one changes a direction of a bias magnetic field to an opposite direction, $\cos\vartheta$ changes its sign and so the helicity density parameter $F$ changes its sign as well [21 – 23, 46]. Since, in this case, the electric- and magnetic-field vectors also change a direction of their rotation, the imaginary part of a vector $\vec{E} \times \vec{H}^*$ is invariant with respect to a direction of a bias magnetic field. From the above analysis it follows that the vacuum regions above and below a quasi-2D ferrite disk, where the helicity density parameter $F$ is nonzero, are also the regions where the imaginary part of a vector $\vec{E} \times \vec{H}^*$ exist as well. For the near-field vacuum areas, localized at an axis of a quasi-2D ferrite disk, we can represent this connection by the following relation:

$$\frac{1}{2}\operatorname{Re}|\vec{E} \cdot \vec{H}^*| = \frac{1}{2}\operatorname{Im}\left[\vec{E} \times (\vec{H})^*\right]_z. \tag{35}$$

So, in a region near a ferrite disk, reactive power flow is accompanied by the helicity factor or, in other words, by the ME-energy density. Our numerical results in the next section clearly show that in a vacuum region where the reactive power flow is observed, the helicity density factor exist as well.

In vacuum regions near a ferrite disk, the helicity density $F$ appears due to the potential electric and potential magnetic fields ($\vec{\nabla} \times \vec{E} = 0$, $\vec{\nabla} \times \vec{H} = 0$). For such a field structure, $\vec{\nabla} \cdot (\vec{E} \times \vec{H}^*) = \vec{H}^* \cdot \vec{\nabla} \times \vec{E} - \vec{E} \cdot \vec{\nabla} \times \vec{H}^* \equiv 0$. At the same time, in a ferrite-material region, where $\vec{\nabla} \times \vec{E} \neq 0$ [21], we have $\vec{\nabla} \cdot (\vec{E} \times \vec{H}^*) = \vec{H}^* \cdot \vec{\nabla} \times \vec{E} \neq 0$. Inside a ferrite disk, one has a localized twist of the field vectors due to the MDM-resonance magnetization motion [19, 21]. It means that a ferrite-material region can be considered as a source region for the vector $\operatorname{Im}\vec{E} \times \vec{H}^*$ observed in a vacuum area. Integration over the disk surfaces and over the ferrite-material-region volume gives

$$\oiint_{S_{disk}} (\operatorname{Im}\vec{E} \times \vec{H}^*) d\vec{S}_{disk} = \int_{V_{disk}} (\vec{\nabla} \cdot \operatorname{Im}\vec{E} \times \vec{H}^*) dV_{disk} \equiv Q_{disk}, \tag{36}$$

where $Q_{disk}$ can be considered as a certain "charge" – a local source of the $\operatorname{Im}\vec{E} \times \vec{H}^*$ vector field in vacuum. We can also conclude that the helicity $\mathcal{H}$ should be related to $Q_{disk}$ by the following integral-form expression:

$$\left|\mathcal{H}^{(+)}\right|+\left|\mathcal{H}^{(-)}\right| \propto Q_{disk}.  \tag{37}$$

## V. NUMERICAL RESULTS: ME-FIELD ACTIVE AND REACTIVE POWER FLOWS AND HELICITY DENSITY

Our numerical studies are based on a commercial finite-element electromagnetic solver (HFSS, Ansoft). In the previous publications [20 – 23, 46, 47], we analyzed numerically the active power flows $\text{Re}\,\vec{E}\times\vec{H}^*$ and the helicity density $F$ for ME fields originated from a ferrite disk with MDM oscillations. A numerical analysis in the present paper is aimed to study the reactive power flows $\text{Im}\,\vec{E}\times\vec{H}^*$ in relation with the active power flows and the helicity density. In the analysis, we use the same disk parameters as in Refs. [20, 21]: the yttrium iron garnet (YIG) disk has a diameter of $D = 3$ mm and the disk thickness is $t = 0.05$ mm; the disk is normally magnetized by a bias magnetic field $H_0 = 4900$ Oe; the saturation magnetization of the ferrite is $4\pi M_s = 1880$ G. A ferrite disk is placed inside a $TE_{10}$-mode rectangular X-band waveguide symmetrically to its walls so that the disk axis is perpendicular to a wide wall of a waveguide. The waveguide walls are made of a perfect electric conductor (PEC). For better understanding the field structures, we assume in our studies that a ferrite disk has very small losses: the linewidth of a ferrite is $\Delta H = 0.1$ Oe. Fig. 2 shows the module of the reflection (the $S_{11}$ scattering-matrix parameter) coefficient. The resonance modes are designated in succession by numbers $n$ = 1, 2, 3… An insert in Fig. 2 shows geometry of the structure. One can see that, starting from the second mode, we have Fano resonances. For every mode number, these coalescent resonances are denoted by single and double primes.

Figs. 3 and 4 show the active and reactive power flows of the ME field near a ferrite disk for the 1$^{st}$ MDM, at the resonance frequency $f_{res} = 8.523$ GHz. We can see that while the active power flow is characterized by the vortex topology, the reactive power flow has a source which is originated from a ferrite disk. The regions of localization of the active and reactive power flows are different. While the active power flow is localized at the disk periphery, the reactive power flow is localized at a central part of the disk. This is well illustrated in Figs. 5 and 6 where intensities of the power flows near a ferrite disk are shown. The helicity density near a ferrite disk for the 1$^{st}$ MDM is shown in Fig. 7. When one compares Fig. 7 (*a*) with Fig. 6 (*b*) and Fig. 7 (b) with Fig. 5 (b), one finds good coincidence between the regions of the reactive-power-flow localization and the helicity-density localization. We have predicted theoretically such a coincidence in the above analysis. It is worth also noting here the regions of localization of the reactive-power-flow and the helicity-density correspond to the region of localization of the potential electric field [21 – 23, 47].

When one changes a direction of a bias field, the active power flow changes its direction as well. Also, the helicity-density factor *F* changes its sign in this case. At the same time, the reactive power flow does not change its direction when the direction of a bias field is changed. The active and reactive power flows are mutually perpendicular. These flows constitute surfaces, which can be considered as deformed versions of the complex planes, i. e. as Riemann surfaces. These behaviors are illustrated in Fig. 8. The helicity-density distribution is related to the angle $\vartheta$ between the spinning electric and magnetic fields. When one moves from the ferrite surfaces, above or below a ferrite disk, one observes reduction of the field amplitudes and also variation of the angle $\vartheta$ between spinning electric and magnetic fields. This angle varies from $0°$ or $180°$ (near the disk surfaces) to $\pm 90°$ (sufficiently far from a ferrite disk). Fig. 9 shows variation of the angle between spinning electric and magnetic fields along the disk axis at different directions of a bias magnetic field. The angle $\vartheta$ gives evidence for a torsion structure of the ME field above and below a ferrite disk. The ME-energy density $\eta$ appears due to such a torsion degree of freedom of the field.

Topological characteristics of the ME fields are quantized quantities. The ME-field topologies appear only at the MDM-resonance frequencies. The ME-field pictures for the 2$^{nd}$ MDM (the peak 2″ in Fig. 2) are shown in Figs. 10 – 14. The main conclusions regarding the power-flow-density and helicity-density distributions expressed for the 1$^{st}$ MDM are also applicable for the 2$^{nd}$ MDM.

In the shown pictures of the power flows and helicity densities, we can observe some small non symmetry, especially for the 2$^{nd}$ MDM. This non symmetry is due to influence of the external microwave radiation which propagates in a waveguide from port 1 to port 2 (see an insert in Fig. 2).

## VI. THE REAL AND IMAGINARY HELICITY DENSITIES OF THE ME FIELDS

As we discussed above, near a ferrite disk with MDM resonances there exist the vacuum regions where the electric and magnetic fields are collinear with each other and are not time-phase shifted. These regions are described by the helicity factor defined by Eq. (18). Now the question arises: Whether there exist also the vacuum near-field regions with collinear the electric and magnetic fields which are time-phase shifted at $90°$? In a case of a positive answer, we should classify the helicity-density factor defined by Eq. (18) (and analyzed in the above study) as a real helicity-density factor:

$$F_{real} = \frac{\omega_{MDM}\varepsilon_0}{4}\text{Im}\left\{i\vec{E}\cdot\vec{B}^*\right\} = \frac{\omega_{MDM}\varepsilon_0}{4}\text{Re}\left\{\vec{E}\cdot\vec{B}^*\right\} = \frac{\omega_{MDM}}{4c^2}\text{Re}\left\{\vec{E}\cdot\vec{H}^*\right\} \qquad (38)$$

and introduce also the notion of an imaginary helicity-density factor:

$$F_{imag} = \frac{\omega_{MDM}\varepsilon_0}{4}\text{Re}\left\{i\vec{E}\cdot\vec{B}^*\right\} = \frac{\omega_{MDM}\varepsilon_0}{4}\text{Im}\left\{\vec{E}\cdot\vec{B}^*\right\} = \frac{\omega_{MDM}}{4c^2}\text{Im}\left\{\vec{E}\cdot\vec{H}^*\right\}. \qquad (39)$$

While the regions of $F_{real}$ are characterized by the potential electric field $\vec{E}_p$, the regions of $F_{imag}$ have the curl-field component $\vec{E}_c$. Based on our numerical analysis, we can show that at the MDM resonances, the near-field regions with imaginary helicity factor exist. The results are represented in Figs. 15 and 16 for the 1$^{st}$ and 2$^{nd}$ MDMs. There is strong difference between the regions of localization of the real and imaginary helicity-density factors. Formally, the imaginary helicity-density factor, espressed by Eq. (39), resembles the optical chirality density [see Eq. (11)]. In our case, however, we do not consider regular electromagnetic fields. Nevertheless, a role of the external electromagnetic fields is evident in Figs. 15 and 16. The pictures are azimuthally nonhomogeneous. Due to the waves propagating in a waveguide, one can see deviation of the topological pictures. Importantly, when one changes a direction of propagation of microwave radiation in a waveguide, the pictures azimuthally turn at angle $180°$.

Could the imaginary helicity-density factor $F_{imag}$ be also related to an imaginary part of the power flow density, as it takes place for a real helicity-density factor $F_{real}$? And, if it is so, what kind of an imaginary part of the power flow density we will have in a case of the imaginary helicity factor $F_{imag}$? To a certain extent, as a preliminary analytical study, we will clarify these questions in the next section.

## VII. THE ME-FIELD HELICITIES AND POWER FLOWS: A VIEW FROM THE MS-WAVE SPECTRAL PROBLEMS

Quantized topological characteristics of the ME fields arise from the MS-wave spectral-problem solutions for MDMs in a quasi-2D ferrite disk [15, 16, 18 – 21]. Detailed analytical studies of the ME-field helicities and power flows based on such MS-wave spectral-problem solutions (being important for our future investigations) are beyond the frames of this work. Nevertheless, a brief examination of this problem is relevant and useful here.

It is worth noting, once again, that the main difference between the active and reactive power flows, as well as between the real and imaginary helicities, appear due to difference of a character of electric fields in the ME-field structure. There are or potential or curl electric fields. In a case of a potential electric field $\vec{E}_p$ (and a potential magnetic field $\vec{H} = -\vec{\nabla}\psi$), we have to solve an integro-differential MS-wave problem. The MS-potential eigenfunction $\psi$ of a certain MDM gives us a potential magnetic field and a dynamical magnetization $\vec{m} = -\bar{\bar{\chi}} \cdot (\vec{\nabla}\psi)$, where $\bar{\bar{\chi}}$ is a magnetic susceptibility tensor [10]. The potential electric field $\vec{E}_p$ in vacuum is calculated by integration over the ferrite-disk region, where the sources – magnetic currents ($\vec{j}^{(m)} = i\omega\vec{m}$) – are given [21]. For a potential electric field $\vec{E}_p$ and a potential magnetic field, one has a reactive power flow and a real helicity density. In the present paper, we obtained these quantities only numerically. Finding of these parameters analytically, based on the integro-differential-problem solutions, is a goal of our future studies.

For a curl electric field $\vec{E}_c$ (and a potential magnetic field $\vec{H} = -\vec{\nabla}\psi$), one has an active power flow and an imaginary helicity density. In this case, the differential-problem analytical solutions for MS waves in a ferrite disk can be easily correlated with our numerical results. A simple manipulation (taking into account that $\vec{\nabla} \cdot \vec{B} = 0$ and $\vec{H} = -\vec{\nabla}\psi$) shows that in this case inside a ferrite region there is:

$$\vec{\nabla} \cdot (\vec{E} \times \vec{H}^*) = \vec{H}^* \cdot \vec{\nabla} \times \vec{E} = i\omega\vec{\nabla}\psi^* \cdot \vec{B} = i\omega\vec{\nabla} \cdot (\psi^*\vec{B}). \qquad (40)$$

With use of the homogeneous boundary conditions, one has the same relation also in a vacuum region near a ferrite disk. Following Ref. [48], one can conclude that this equation gives

$$\vec{E} \times \vec{H}^* = i\omega\psi^*\vec{B}. \qquad (41)$$

A simple analysis of the energy balance equation for monochromatic MS waves in a magnetic medium with small losses [18] shows that, definitely, $\text{Re}(i\omega\psi^*\vec{B})$ is a real power flow density. So, for MS waves the quantity $\text{Re}(\vec{E} \times \vec{H}^*)$ appears as real power flow density as well. Importantly, despite the fact that expression $\text{Re}(\vec{E} \times \vec{H}^*)$ looks like the Poynting vector, the MS-wave real power flow density cannot be interpreted as the EM-wave real power flow density. The Poynting vector is obtained for EM radiation which is described by the *two curl operator* Maxwell equations for the electric and magnetic fields. This is not the case of the MS waves, where we have *potential magnetic* and *curl electric* fields. Since Eqs. (40) and (41) are relevant only for a curl electric field in a vacuum region near a ferrite disk, there is no connection of these equations to the quantity $\text{Im}(\vec{E} \times \vec{H}^*)$ which is related to a potential electric field in vacuum.

We consider now the spectral problem of MDM oscillations in a quasi-2D ferrite disk. For a coordinate system shown in Fig. 1, the MS-potential wave function $\psi$ for a certain mode $n$ can be represented as [16, 18, 19, 47]

$$\psi_n = C_n \xi_n(z) \tilde{\varphi}_n(r,\theta), \tag{42}$$

where $\xi_n(z)$ is an amplitude factor, $C_n$ is a dimensional coefficient, and $\tilde{\varphi}_n(r,\theta)$ is a dimensionless membrane function. The function $\xi_n(z)$ describes a standing wave along $z$ axis. For membrane function, we have $\tilde{\varphi}_n = \tilde{\varphi}_n(r)\tilde{\varphi}_n(\theta)$, where $\tilde{\varphi}_n(r)$ is a Bessel function and $\tilde{\varphi}_n(\theta) \sim e^{-i\nu_n \theta}$. Here $\nu_n$ is an integer ($n$ is a number of radial variations for a given azimuth number $\nu$). For magnetic flux density ($\vec{B}_n = -\vec{\mu} \cdot \vec{\nabla} \psi_n$), we have

$$\vec{B}_n = (B_n)_z \vec{e}_z + \tilde{B}_n \vec{e}_\perp, \tag{43}$$

where

$$(B_n)_z = -C_n \frac{\partial \xi_n(z)}{\partial z} \tilde{\varphi}_n(r,\theta) \tag{44}$$

and

$$\tilde{B}_n = -C_n \xi_n(z) \left[ \vec{\mu}_\perp \cdot \vec{\nabla}_\perp \tilde{\varphi}_n(r,\theta) \right] \cdot \vec{e}_\perp. \tag{45}$$

Here subscript $\perp$ corresponds to the transversal (with respect to $z$ axis) vector and tensor components.

The real power flow density for a MDM $n$ is expressed as

$$\vec{p}_n = \text{Re}(i\omega \psi_n^* \vec{B}_n) = \frac{i\omega}{2} (\psi_n^* \vec{B}_n - \psi_n \vec{B}_n^*). \tag{46}$$

It is easily to show [47] that $(p_n)_r = (p_n)_z = 0$. The only non-zero component of the real power flow density is the azimuth component, which is expressed as [47]:

$$\begin{aligned}(p_n(r,z))_\theta &= \frac{i\omega}{2} C_n^2 \left(\xi(z)\right)^2 \left[ -\mu \frac{1}{r} \left( \tilde{\varphi}_n^* \frac{\partial \tilde{\varphi}_n}{\partial \theta} - \tilde{\varphi}_n \frac{\partial \tilde{\varphi}_n^*}{\partial \theta} \right) + i\mu_a \left( \tilde{\varphi}_n^* \frac{\partial \tilde{\varphi}_n}{\partial r} + \tilde{\varphi}_n \frac{\partial \tilde{\varphi}_n^*}{\partial r} \right) \right] \\ &= -\omega\, C_n^2 \left(\xi_n(z)\right)^2 \left[ \frac{\mu}{r} \nu_n \left| \tilde{\varphi}_n(r,\theta) \right|^2 + \frac{1}{2}\mu_a \frac{\partial \left|\tilde{\varphi}_n(r,\theta)\right|^2}{\partial r} \right],\end{aligned} \qquad (47)$$

where $\mu$ and $\mu_a$ are, respectively, the diagonal and off-diagonal components of the tensor $\vec{\mu}$.

The quantities $(p_n(r,z))_\theta$, circulating around a circle $2\pi r$, are the MDM power-flow-density vortices with cores at the disk center. At a vortex center the amplitude of $(p_n)_\theta$ is equal to zero. It follows from Eq. (47) that for a given mode number $n$, characterizing by a certain function $\tilde{\varphi}_n(r)$, there are oppositely directed power-flow vortices for different signs of the azimuth number $\nu_n$. In a vacuum region, outside a ferrite disk we have from Eq. (47)

$$(p_n(r,z))_\theta = -\omega\, C_n^2 \frac{1}{r} \nu_n \left(\xi_n(z)\right)^2 \left|\tilde{\varphi}_n(r,\theta)\right|^2. \qquad (48)$$

It is clear that $\vec{\nabla} \cdot \vec{p}_n = 0$, both inside and outside a ferrite disk. The analytical results for MDM power-flow-density vortices, obtained based on Eqs. (47), (48) are in good correlation with the numerical results. It is worth noting, once again, that following Eqs. (40), (41), analytical solutions for $(p_n)_\theta$ in vacuum are obtained for the curl electric fields.

Let us extend our analysis by introduction also an imaginary MS-wave power flow. It means that we will consider now the complex quantity

$$\vec{s} = \vec{p} + i\vec{q}, \qquad (49)$$

where $\vec{p} = \mathrm{Re}\left(i\omega\psi^*\vec{B}\right) = \frac{i\omega}{2}\left(\psi^*\vec{B} - \psi\vec{B}^*\right)$ and $\vec{q} = \mathrm{Im}\left(i\omega\psi^*\vec{B}\right) = \frac{\omega}{2}\left(\psi^*\vec{B} + \psi\vec{B}^*\right)$. The physical meaning of the quantity $\mathrm{Re}\left(i\omega\psi^*\vec{B}\right)$ is clear from the above analysis. Now the question arises: What does it mean the quantity $\mathrm{Im}\left(i\omega\psi^*\vec{B}\right)$?

For the imaginary power flow density of a MDM $n$, one has

$$\vec{q}_n = \frac{\omega}{2}\left(\psi_n^*\vec{B}_n + \psi_n\vec{B}_n^*\right). \qquad (50)$$

In this case, we have $(q_n)_\theta = 0$, while $(q_n)_r \neq 0$ and $(q_n)_z \neq 0$. For $(q_n)_r$, we obtain:

$$\begin{aligned}(q_n(z))_r &= -\frac{\omega}{2}C_n^2(\xi(z))^2\left[\mu\left(\tilde{\varphi}_n^*\frac{\partial\tilde{\varphi}_n}{\partial r}+\tilde{\varphi}_n\frac{\partial\tilde{\varphi}_n^*}{\partial r}\right)+i\mu_a\frac{1}{r}\left(\tilde{\varphi}_n^*\frac{\partial\tilde{\varphi}_n}{\partial\theta}-\tilde{\varphi}_n\frac{\partial\tilde{\varphi}_n^*}{\partial\theta}\right)\right]\\ &= -\omega C_n^2(\xi(z))^2\left[\frac{1}{2}\mu\frac{\partial|\tilde{\varphi}_n(r)|^2}{\partial r}-\mu_a\frac{\nu_n}{r}|\tilde{\varphi}_n(r)|^2\right].\end{aligned} \qquad (51)$$

Outside a ferrite disk we have

$$(q_n(\theta,z))_r = -\frac{\omega}{2}C_n^2(\xi(z))^2\frac{\partial|\tilde{\varphi}_n(r)|^2}{\partial r}. \qquad (52)$$

For $(q_n)_z$, we obtain:

$$(q_n)_z = -\omega C_n^2\xi_n(z)\frac{\partial\xi_n(z)}{\partial z}|\tilde{\varphi}_n(r,\theta)|^2. \qquad (53)$$

Since the quantity $q_n$ is related to a curl electric field in a vacuum region, it does not describe analytically the obtained above quantity $\text{Im}(\vec{E}\times\vec{H}^*)$, which is related to a potential electric field in vacuum. We can suppose, however, that the analytical quantity $q_n$ is related to the imaginary helicity-density factor $F_{imag}$, which corresponds to a curl electric field in vacuum. Study of this relationship is a goal for our future analysis. In solving this problem we should take into account also a role of the external microwave radiation. The azimuth nonhomogeneity of $F_{imag}$, shown in Figs. 15 and 16, is due to these external RF fields.

## VIII. DISCUSSION AND CONCLUSION

The dual symmetry between electric and magnetic fields underlies the conservation of energy and momentum for electromagnetic fields and also the conservation of optical (electromagnetic) helicity. Symmetry properties of the ME fields are different from symmetry properties of free-space electromagnetic fields. The near fields originated from small ferrite particles with MDM oscillations are the fields with the electric and magnetic components, but with broken dual (electric-magnetic) symmetry. These fields – the ME fields – have topological properties different from such properties of electromagnetic fields. In this paper, we showed that topological properties of ME fields are presented by a very complicated picture. There are real and imaginary ME power flows and real and imaginary helicity densities. These quantities are related to the analytically derived MS-potential quadratic relations.

In classical electrodynamics, the imaginary part of the Poynting vector gives information about evanescent (i.e. non-propagating) fields. For a plane wave, the imaginary part of the Poynting vector is zero. When, however, the Poynting vector is calculated from the fields in some below-cut-off waveguiding structures, a non-zero imaginary part of the Poynting vector can be found. This indicates that there are resonant, non-propagating fields at that location. In regular lossless structures, or real or imaginary Poynting vector exists. In a case of the ME-field structure, situation with the power flows is completely different. In this paper we showed that in the vacuum area near a ferrite disk with MDM resonances, one can observe both active ($\text{Re}\,\vec{E}\times\vec{H}^*$) and reactive ($\text{Im}\,\vec{E}\times\vec{H}^*$) power flows. While active power flows are characterized by the vortex topology, reactive power flows are source originated from a ferrite disk. These flows constitute surfaces, which can be considered as deformed versions of the complex planes, i. e. as Riemann surfaces. The regions of localization of the active and reactive power flows are different. When the active power flow is localized at the disk periphery, the reactive power flow is localized at a central part of the disk. There is also an evident coincidence between the regions of the reactive-power-flow localization and the real-helicity-density localization. We showed that in an area of these localizations, there is variation of the angle between spinning electric and magnetic fields along the disk axis. This angle gives evidence for a torsion structure of the ME field above and below a ferrite disk. The ME-energy density $\eta$ appears due to the torsion degree of freedom of the field. The near-field vacuum regions where the electric and magnetic fields are collinear with each other and are not time-phase shifted are the regions with the real helicity-density factor. We showed that together with such regions there are also the vacuum near-field regions where the electric and magnetic fields are collinear with each other but are time-phase shifted at $90°$. In the last case, we are talking about the imaginary helicity-density factor. The observed quantized topological characteristics of the ME fields arise from the MS-wave spectral-problem solutions for MDMs in a quasi-2D ferrite disk. A preliminary analytical examination of the ME-field helicities and power flows based on such MS-wave spectral-problem solutions, made in the paper, shows important relationship between the numerical and analytical results.

The shown topological properties of the ME fields can be useful for novel near- and far-field microwave applications. Strongly localized ME fields, having both the real and imaginary helicity parameters, open unique perspective for sensitive microwave probing of structural characteristics of chemical and biological objects. The presence of a biological sample with chiral properties will necessarily alter the near-field distribution which in turn will change the spectral characteristics of a MDM ferrite disk. Regarding the far-field applications, some interesting questions can be expressed here. The active ($\text{Re}\,\vec{E}\times\vec{H}^*$) and reactive ($\text{Im}\,\vec{E}\times\vec{H}^*$) power flows exist together and are mutually perpendicular one to another. We can also see that the sources of the reactive power flows are at the centers of active-power-flow vortices. With these topological properties, the following question arises: Can we observe free space transportation of reactive energy by an active power flow? In Ref. [22], we showed the effect of long-distance interaction between two MDM ferrite disks. We have found that the split-resonance response for coupled MDM particles is weakly dependent on distances between disks. Such coupling was clearly observed even at extremely long distances between the disks. One of the proper explanations of the effect in Ref. [22] can, probably, appear based on our studies in the present paper. Because of a joint structure of the reactive and active power flows and transportation of reactive energy by an active power flow, we may have a long-distance (reactive) coupling between two MDM ferrite disks. This is a quasistatic phase coupling between two MDM disks, which very weakly depends on a distance between the disks. It can be supposed that in Ref. [22], we observed such an effect of a quasistatic MDM coupling.

[47] M. Sigalov, E. O. Kamenetskii, and R. Shavit, J. Phys.: Condens. Matter **21**, 016003 (2009).
[48] D. D. Stancil, *Theory of Magnetostatic Waves* (Springer-Verlag, New York, 1992).20

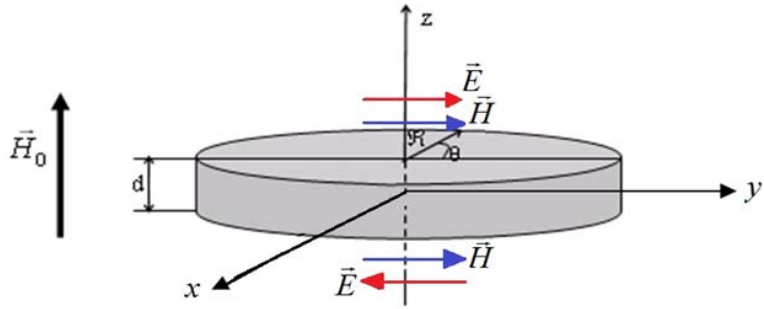

Fig. 1. Spinning electric- and magnetic-field vectors in vacuum regions above and below a MDM-resonance ferrite disk.

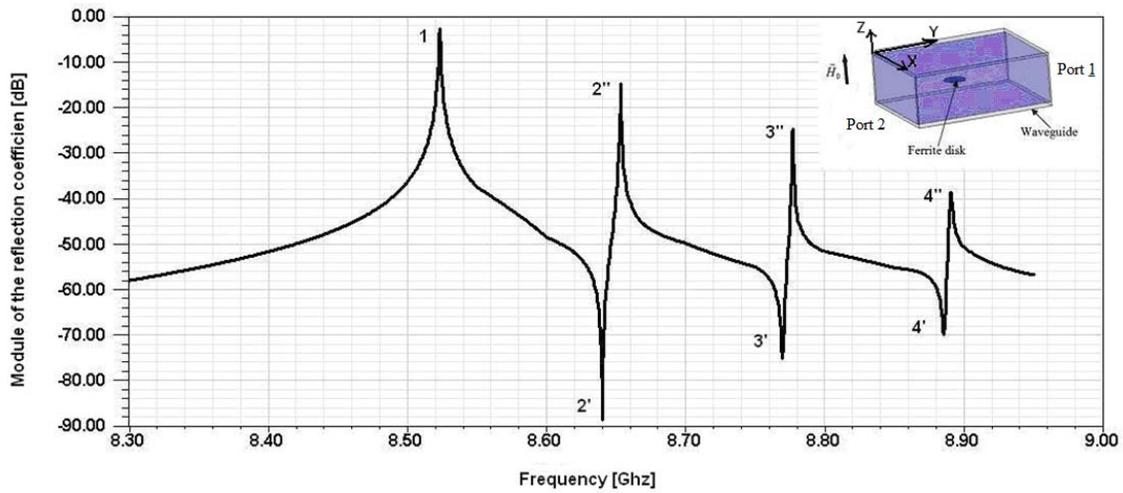

Fig. 2. Frequency characteristics of a module of the reflection coefficient for a rectangular waveguide with an enclosed thin-film ferrite disk. The resonance modes are designated in succession by numbers n = 1, 2, 3… The coalescent (Fano-type) resonances are denoted by single and double primes. An insert shows geometry of a structure.

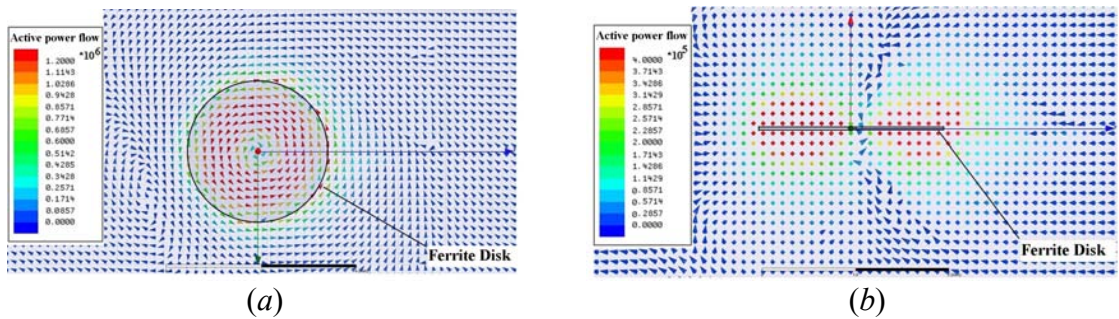

(a)  (b)

Fig. 3. Active power flow density near a ferrite disk for the 1$^{st}$ MDM ($f_{res} = 8.523$ GHz). A bias magnetic field is upward directed along $z$ axis. (a) A view on a vacuum plane parallel to the ferrite-disk plane and at distance 50 microns above a disk, (b) a view on a cross-section plane perpendicular to the ferrite-disk plane.

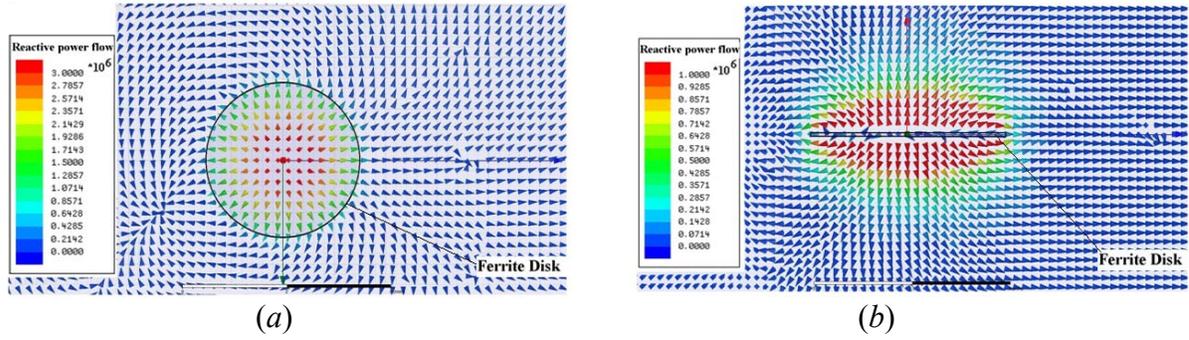

Fig. 4. Reactive power flow density near a ferrite disk for the 1$^{st}$ MDM ($f_{res} = 8.523$ GHz). (*a*) A view on a vacuum plane parallel to the ferrite-disk plane and at distance 50 microns above a disk, (*b*) a view on a cross-section plane perpendicular to the ferrite-disk plane.

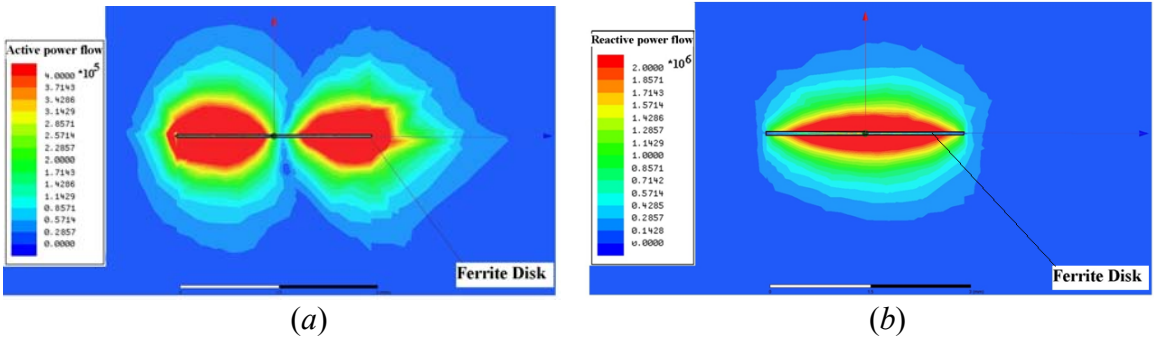

Fig. 5. Intensity of the power flows near a ferrite disk for the 1$^{st}$ MDM on a cross-section plane perpendicular to the ferrite-disk plane. (*a*) Active power flow; (*b*) reactive power flow.

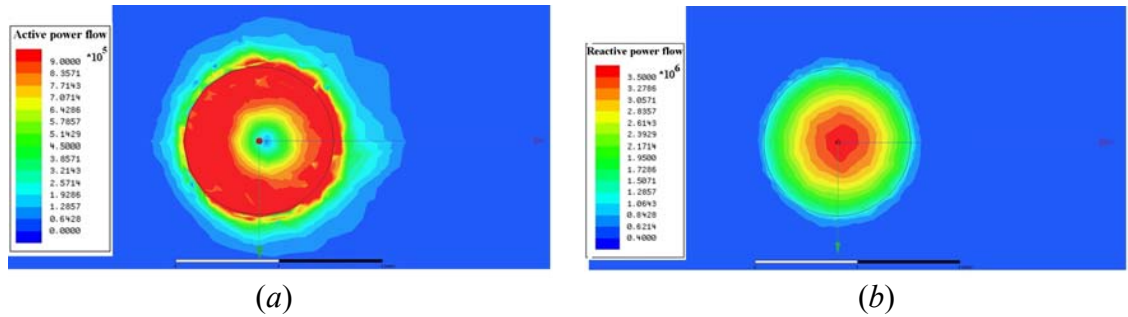

Fig. 6. Intensity of the power flows near a ferrite disk for the 1$^{st}$ MDM ) on a vacuum plane parallel to the ferrite-disk plane and at distance 50 microns above a disk. (*a*) Active power flow; (*b*) reactive power flow.

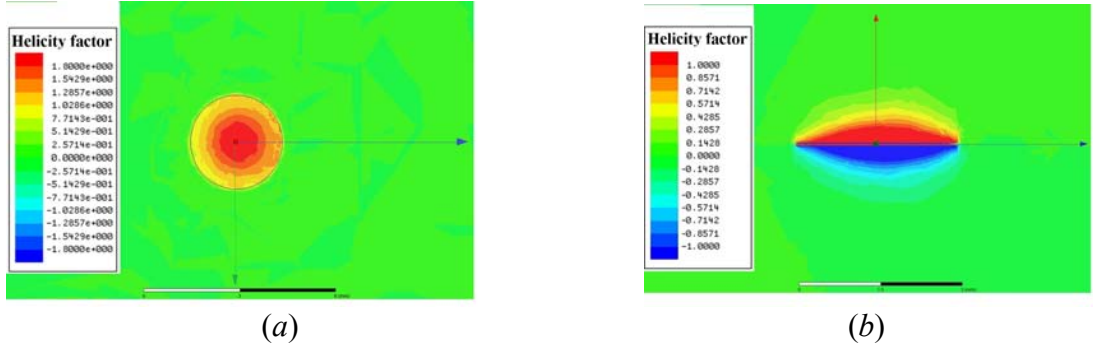

(a)                          (b)

Fig. 7. The helicity density near a ferrite disk for the 1$^{st}$ MDM. A bias magnetic field is upward directed along *z* axis. (*a*) A view on a vacuum plane parallel to the ferrite-disk plane and at distance 50 microns above a disk, (*b*) a view on a cross-section plane perpendicular to the ferrite-disk plane.

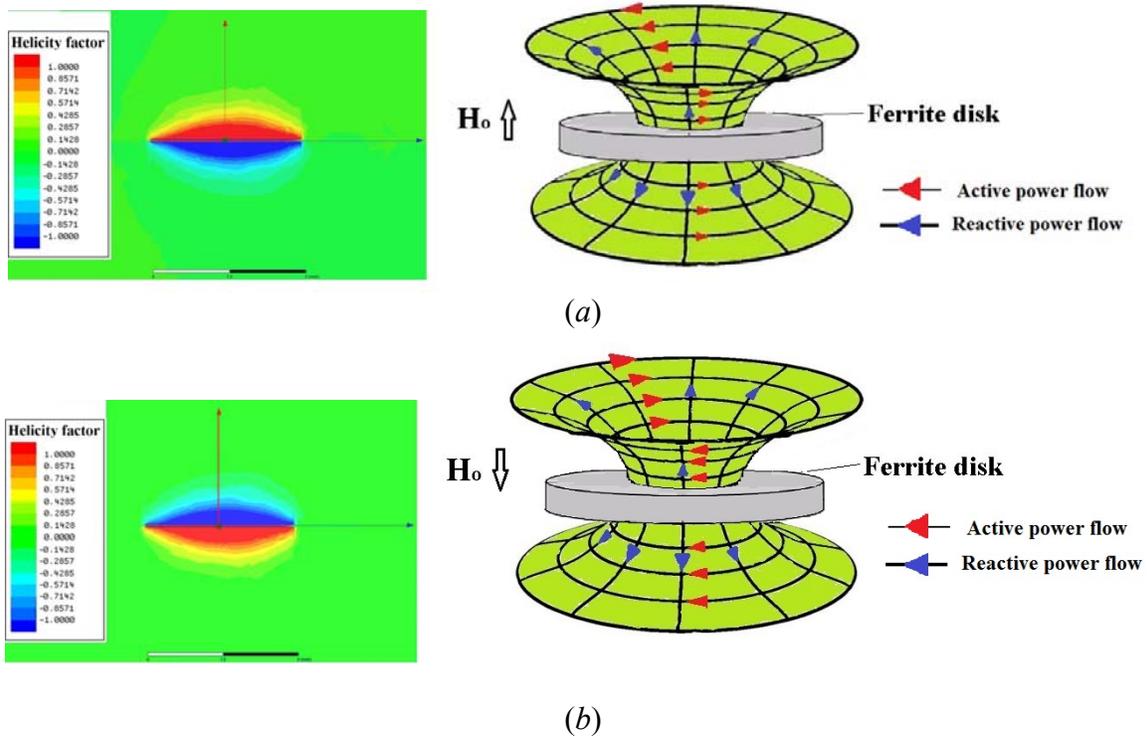

(*a*)

(*b*)

Fig. 8. The helicity density and active and reactive power flows for the 1$^{st}$ MDM ($f_{res} = 8.523$ GHz). (*a*) An upward directed bias magnetic field; (*b*) a downward directed bias magnetic field. The active and reactive power flows are mutually perpendicular. These flows constitute surfaces, which can be considered as deformed versions of the complex planes, i. e. as Riemann surfaces. When one changes a direction of a bias field, the active power flow changes its direction as well. Also, the helicity-density factor *F* changes its sign in this case. At the same time, the reactive power flow does not change its direction when the direction of a bias field is changed.

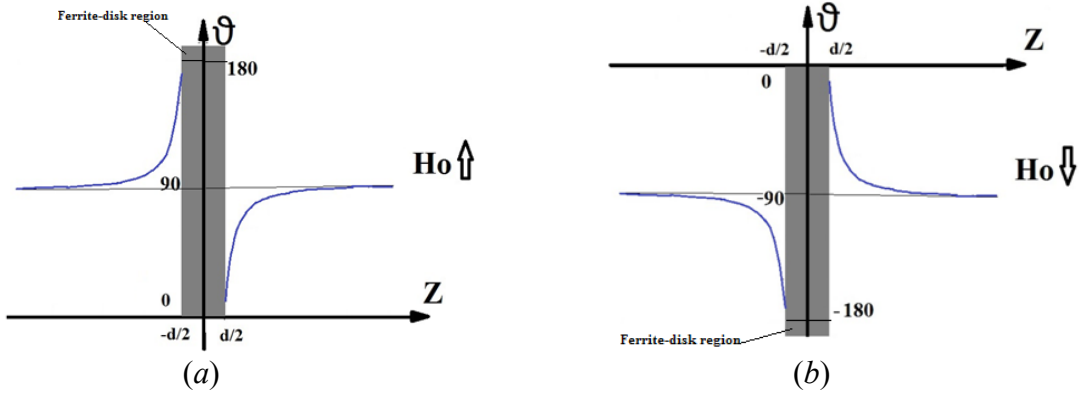

Fig. 9. Variation of the angle between spinning electric and magnetic fields along the disk axis for the 1$^{st}$ MDM. This angle gives evidence for a torsion structure of the ME field above and below a ferrite disk. The ME-energy density $\eta$ appears due to the torsion degree of freedom of the field. (*a*) Angle $\vartheta$ for an upward directed bias magnetic field; (*b*) angle $\vartheta$ for a downward directed bias magnetic field.

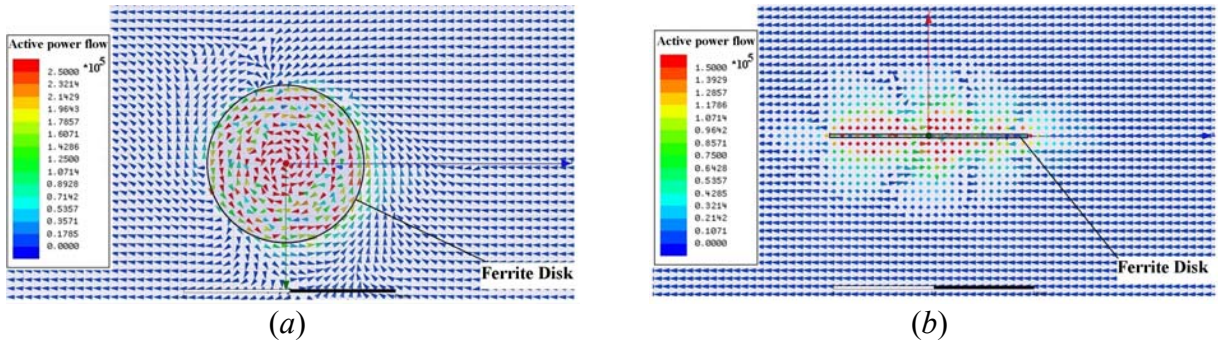

Fig. 10. Active power flow density near a ferrite disk for the 2$^{nd}$ MDM ($f_{res} = 8.653$ GHz). A bias magnetic field is upward directed along *z* axis. (*a*) A view on a vacuum plane parallel to the ferrite-disk plane and at distance 50 microns above a disk, (*b*) a view on a cross-section plane perpendicular to the ferrite-disk plane.

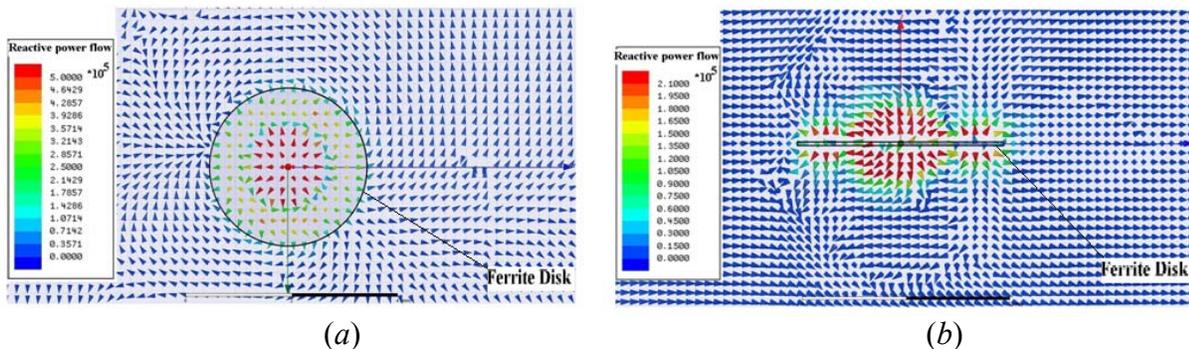

Fig. 11. Reactive power flow density near a ferrite disk for the 2$^{nd}$ MDM ($f_{res} = 8.653$ GHz). (*a*) A view on a vacuum plane parallel to the ferrite-disk plane and at distance 50 microns above a disk, (*b*) a view on a cross-section plane perpendicular to the ferrite-disk plane.

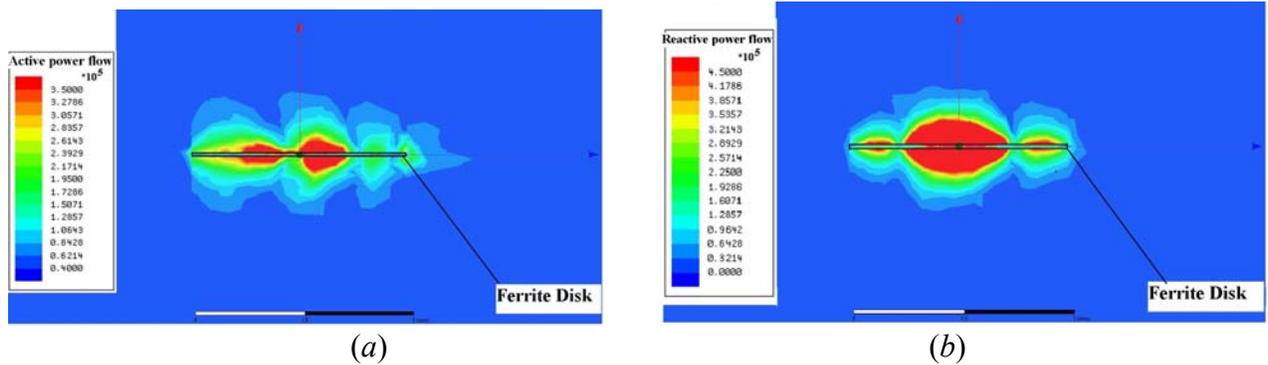

Fig. 12. Intensity of the power flows near a ferrite disk for the 2nd MDM on a cross-section plane perpendicular to the ferrite-disk plane. (*a*) Active power flow; (*b*) reactive power flow.

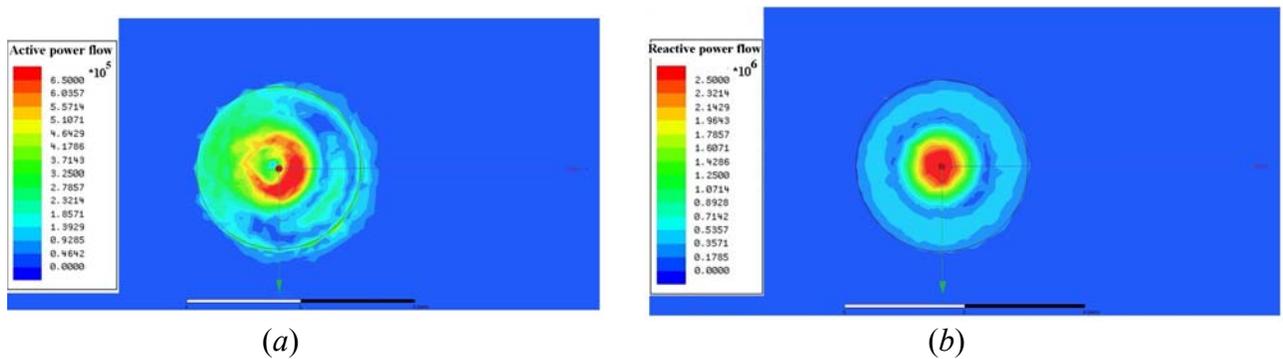

Fig. 13. Intensity of the power flows near a ferrite disk for the 2nd MDM on a vacuum plane parallel to the ferrite-disk plane and at distance 50 microns above a disk. (*a*) Active power flow; (*b*) reactive power flow.

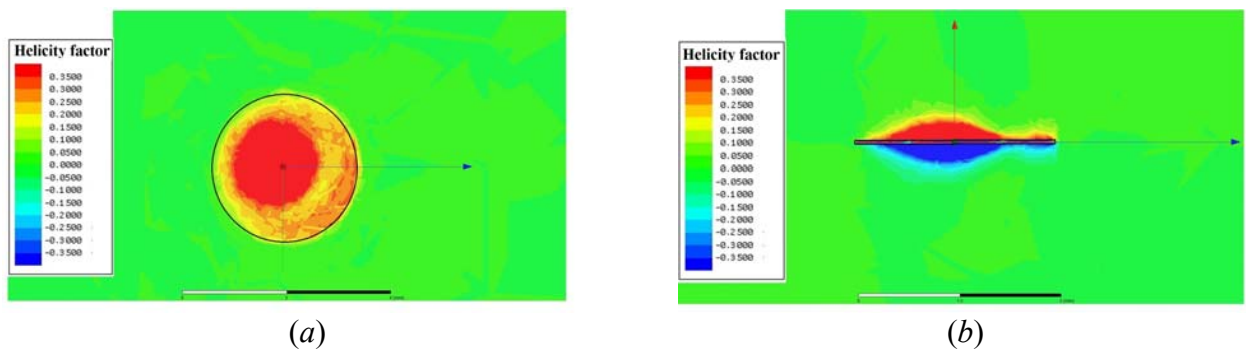

Fig. 14. The helicity density near a ferrite disk for the 2nd MDM. A bias magnetic field is upward directed along *z* axis. (*a*) A view on a plane parallel to the ferrite-disk plane and at distance 50 microns above a disk; (*b*) a view on a cross-section plane perpendicular to the ferrite-disk plane.

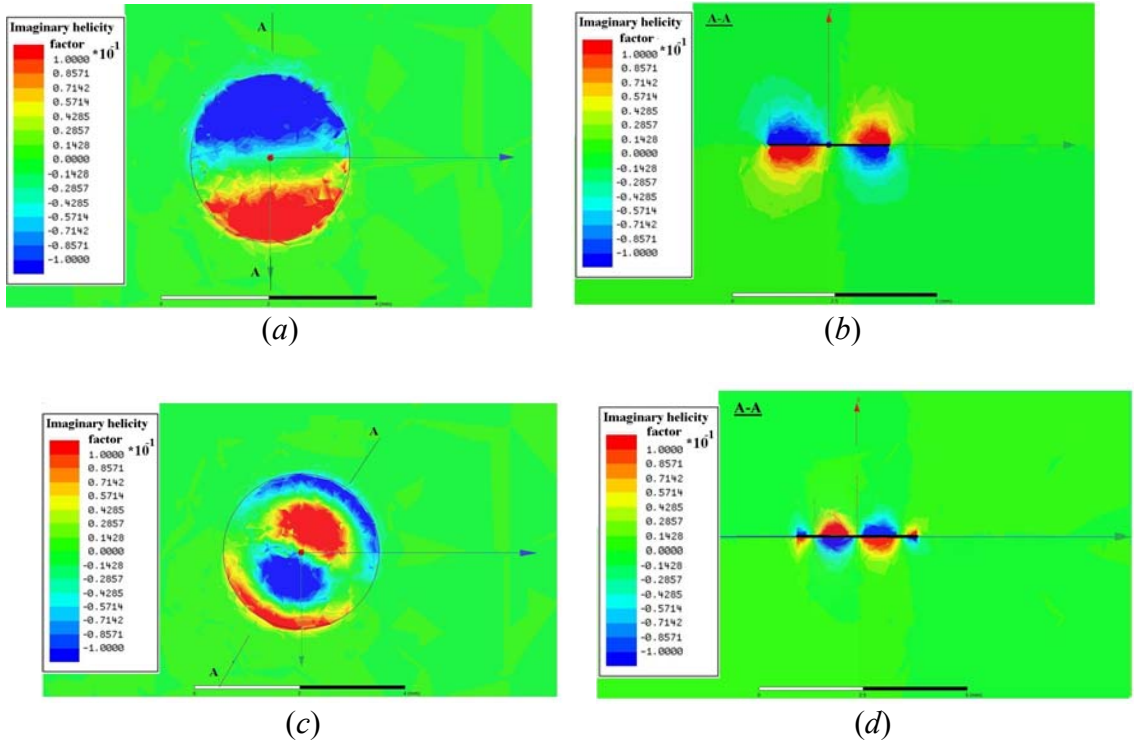

Fig. 15. The imaginary helicity density near a ferrite disk for the 1st [(*a*) and (*b*)] and the 2nd [(*c*) and (*d*)] MDMs. A bias magnetic field is upward directed along *z* axis. Microwave radiation propagates in a waveguide from port 1 to port 2 (see an insert in Fig. 2). (*a*), (*c*) A view on a vacuum plane parallel to the ferrite-disk plane and at distance 50 microns above a disk; (*b*), (*d*) a view on a cross-section A-A perpendicular to the ferrite-disk plane.

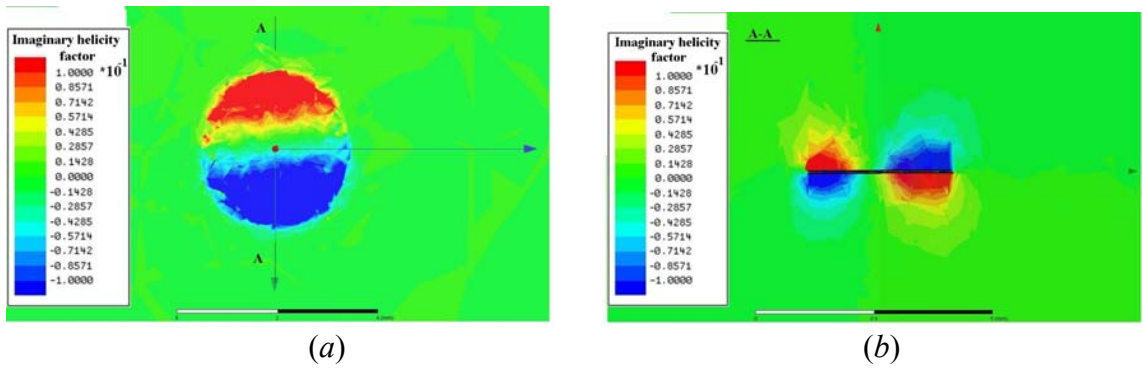

(*a*)          (*b*)

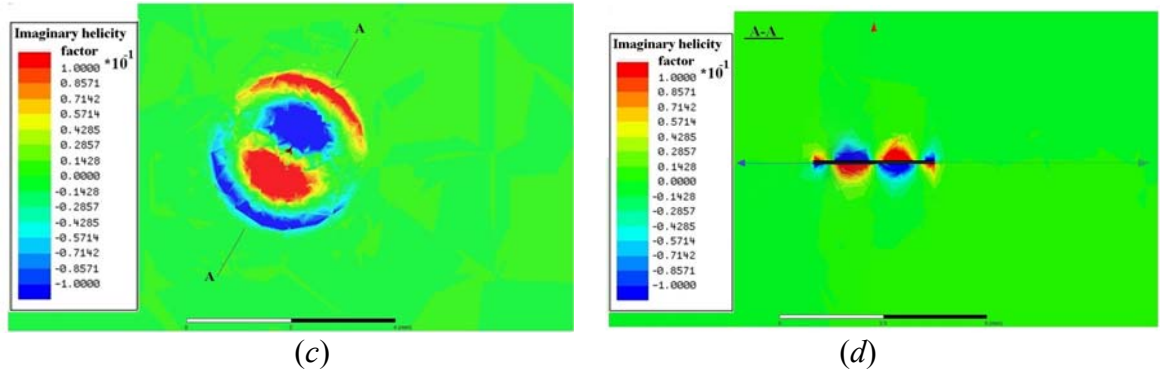

Fig. 16. The imaginary helicity density near a ferrite disk for the 1st [(*a*) and (*b*)] and the 2nd [(*c*) and (*d*)] MDMs. A bias magnetic field is upward directed along *z* axis. Microwave radiation propagates in a waveguide from port 2 to port 1 (see an insert in Fig. 2). (*a*), (*c*) A view on a vacuum plane parallel to the ferrite-disk plane and at distance 50 microns above a disk; (*b*), (*d*) a view on a cross-section A-A perpendicular to the ferrite-disk plane.